# MOSP: A User-interface Package for Simulating Metal Nanoparticle Structure and Reactivity under Operando Conditions

Lei Ying[1,2], Beien Zhu[1,3*], Yi Gao[1,3,4*]

[1]Key Laboratory of Interfacial Physics and Technology, Shanghai Institute of Applied Physics, Chinese Academy of Sciences, Shanghai 201800, China.

[2]University of Chinese Academy of Sciences, Beijing 100049, China.

[3]Phonon Science Research Center for Carbon Dioxide, Shanghai Advanced Research Institute, Chinese Academy of Sciences, Shanghai 201210, China.

[4]Key Laboratory of Low-Carbon Conversion Science & Engineering, Shanghai Advanced Research Institute, Chinese Academy of Sciences, Shanghai 201210, China

*Corresponding author. Email: gaoyi@sari.ac.cn (Y.G.); zhube@sari.ac.cn (B.Z.).

**Abstract**:

Structures of metal nanoparticles (NPs) significantly influence their catalytic reactivities. Recent *in situ* experimental observations of dramatic structural changes in NPs have underscored the need to establish a dynamic structure-property relationship that accounts for the reconstruction of NPs in reactive environments. Here, we present the `MOSP`, a free and open-source graphical user interface (GUI) package designed to simulate the structure and reactivity of metal NPs under operando conditions. `MOSP` integrates two models: the multiscale structure reconstruction (MSR) model predicting equilibrium metal NP structures under specific reaction conditions and the kinetic Monte Carlo (KMC) model simulating the reaction dynamics. This combination allows for the exploration of the dynamic structure-property relationships of NPs. `MOSP` enhances user accessibility through its intuitive GUI, facilitating easy input, post-processing, and visualization of simulation data. This article is the release note of `MOSP`, focusing on its implementation and functionality.

# I. INTRODUCTION

Metal nanoparticles (NPs) play an essential role in heterogeneous catalysis. The catalytic properties of NPs are highly correlated with their structures, which directly influence the number and assembly of the active sites.[1-5] This highlights the importance of investigating structure-activity relationships for the rational design of efficient catalysts. This relationship can be theoretically explored through first-principle-based kinetic simulations, such as mean-field microkinetic modeling (MKM) and kinetic Monte Carlo (KMC) simulation.[6-8] These methods, based on density functional theory (DFT) calculated energy profiles, enable the prediction of catalytic behavior of NPs with given structures under specific operating conditions (e.g. temperature and pressure) without the reliance on experimental data.

In previous modeling, model structures that does not dependent on reaction conditions were widely used. However, recent *in situ* experiments have revealed that the structure of metal NPs can dynamically change in response to changing reaction conditions.[9-13] This challenges traditional kinetic simulations based on fixed structures, as they do not account for reaction-condition-induced shape change of the metal NPs. To address this, a multiscale structure reconstruction (MSR) model has been developed in our previous works.[14-16] This model quantitatively predicts the equilibrium structure of metal NPs under reaction conditions from a thermodynamic perspective and has been validated against several experimental observations.[16-20] By combining MSR with reaction KMC, we can explore the dynamic structure–activity relationship under reaction conditions can be achieved.[21-23]

Herein, we present the Multi-scale Operando Simulation Package (`MOSP`), an open-source graphical user interface (GUI) application written in `python`. `MOSP` comprises two main modules: MSR for constructing equilibrium NP structures under operando conditions and KMC for simulating catalytic properties. The GUI enhances the accessibility of `MOSP` through intuitive user inputs. Interactive data post-processing and visualization features have also been implemented. The implementation of `MOSP` is outlined, and its capabilities are demonstrated through two illustrative examples: CO oxidation over Pt NPs, and water-gas shift (WGS) reaction over Cu NPs.

# II. THEORETICAL BACKGROUND
## A. MSR model

According to the Wulff theory, the equilibrium shape of a NP is determined by the surface tension of each facet ($\gamma_{hkl}$).[24] When the NP is exposed to the reaction atmosphere, $\gamma_{hkl}$ is modified to the interface tension $\gamma_{hkl}^{int}$.[14-16] In a reaction atmosphere comprising $n$ types of gases, the coverages and adsorption energies of these gases can be represented by a vector of length $n$ on each facet ($hkl$), denoted as $\boldsymbol{\theta}^{hkl}$ and $\boldsymbol{E}^{hkl}$, respectively. The lateral interactions among adsorbates can be expressed as an $n \times n$ symmetric matrix, denoted as $\boldsymbol{w}^{hkl}$, where each element $w_{ij}^{hkl}$ indicating the lateral interaction between the $i^{th}$ and $j^{th}$ species. For example, under a CO and O₂ (dissociative adsorption) binary gas environment, the lateral interactions $\boldsymbol{w}$ can be expressed as:

$$\boldsymbol{w} = \begin{bmatrix} w_{CO-CO} & w_{CO-O} \\ w_{CO-O} & w_{O-O} \end{bmatrix} \qquad (1)$$

The interface tension $\gamma_{hkl}^{int}$ under the reaction atmosphere could be obtained as follows:

$$\gamma_{hkl}^{int} = \gamma_{hkl} + \frac{1}{A_{at}^{hkl}} \left( \boldsymbol{\theta}^{hkl} \cdot (\boldsymbol{E}^{hkl} - z^{hkl} \boldsymbol{\theta}^{hkl} \cdot \boldsymbol{w}^{hkl}) \right) \qquad (2)$$

Here, $A_{at}^{hkl}$ is the surface area per atom on this facet, and $z^{hkl}$ is the number of the surface nearest neighbors. The coverages $\boldsymbol{\theta}^{hkl}$ are obtained by solving the Fowler-Guggenheim (F-G) adsorption isotherm.[25] For the $i^{th}$ gas, at equilibrium, the coverage $\theta_i^{hkl}$ conforms to:

$$\frac{\theta_i^{hkl}}{1 - \sum_{j=1}^{n} \theta_j^{hkl}} = (P_i K_i^{hkl})^{\frac{1}{\alpha_i}} \exp\left( \frac{\alpha_i z^{hkl} \sum_{j=1}^{n} w_{ij}^{hkl} \theta_j^{hkl}}{k_B T} \right) \qquad (3)$$

where $k_B$ is the Boltzmann constant, $T$ is the temperature, $P_i$ is the partial pressure of the gas, and $\alpha_i$ is a parameter that depends on the form of adsorption, which is 1 for associative adsorption, and 2 for dissociative adsorption.[26] The equilibrium constant $K_i^{hkl}$ is defined as

$$K_i^{hkl} = \exp(-\frac{\Delta G}{k_B T}) = \exp\left( -\frac{\alpha_i E_i^{hkl} - T(\alpha_i S_{ads,i}^{hkl} - S_{gas,i})}{k_B T} \right) \qquad (4)$$

where $S_{gas,i}$ and $S_{ads,i}^{hkl}$ are the entropies of the gas before and after adsorption on the facet ($hkl$), respectively. $S_{gas,i}$ is evaluated from the standard entropy $S_{gas,i}^0$ (Eq. 5), which can be obtained from the NIST-JANANF thermochemical tables[27].

$$S_{gas,i} = S_{gas,i}^0 - k_B ln\left(\frac{P_i}{P^0}\right) \qquad (5)$$

where $P^0$ is 1 bar.

Furthermore, considering the finite-size effect of metal NPs,[5, 28] MSR model is not recommended for the construction of particles smaller than 3 nm in size.

## B. Kinetic Monte Carlo approach

In a kinetic Monte Carlo (KMC) simulation, the state of a gas-surface catalyzed system is described by defining a surface configuration.[29-31] The potential energy surface (PES) can be coarse-grained into a set of energy basins, which are separated by chemical transitions between surface configurations.[31, 32] The master equation (ME) describing the time evolution of this system can be derived from first principles.[29, 31] Specifically,

$$\frac{dP_\alpha}{dt} = \sum_\beta [W_{\alpha \leftarrow \beta} P_\beta - W_{\beta \leftarrow \alpha} P_\alpha] \tag{6}$$

Here, $P_\alpha$ ($P_\beta$) is the probability that the system is in surface configuration $\alpha$ ($\beta$), while $W_{\alpha \leftarrow \beta}$ ($W_{\beta \leftarrow \alpha}$) is the transition rate from $\beta$ to $\alpha$ ($\alpha$ to $\beta$). The ME can be solved numerically by KMC simulations,[30] implemented by serval algorithms, such as the First Reaction Method[33, 34] (FRM), the Direct Method[33, 35] (DM), and the Random Selection Method[36] (RSM).

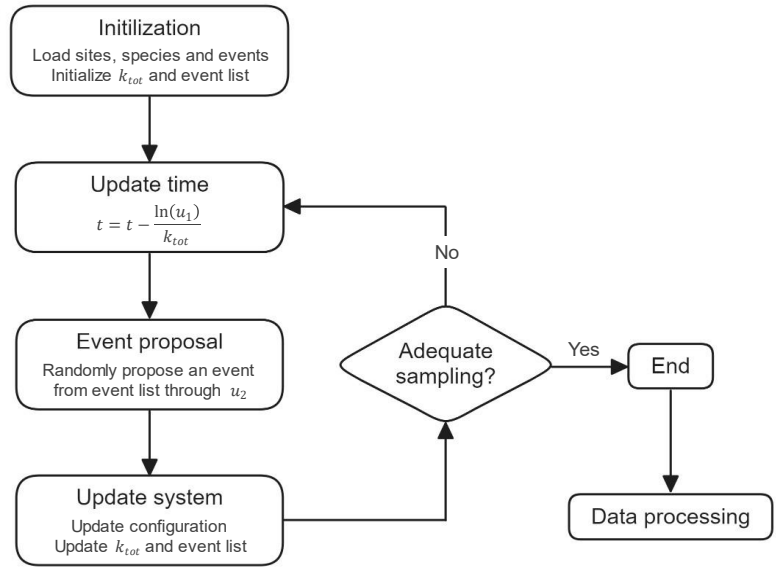

**Fig. 1.** The DM kinetic Monte Carlo algorithm

In MOSP, the surface configuration is represented by a set of occupation values assigned to each site, with the value specifying the adsorbed species at this site (0 for an empty site). Transitions between configurations proceed through various surface events (i.e. adsorption, desorption, diffusion, and reaction). Each event changes the occupations of one or two sites. The Gillespie's DM algorithm[33, 35] is adopted. At each

KMC step, the time increment is calculated based on the total rate constant of all available events ($k_{tot} = \sum k_{event}$):

$$\Delta t = -\frac{\ln(u_1)}{k_{tot}} \tag{7}$$

where $u_1$ is a uniformly distributed random number in the range (0, 1]. One of the available events is chosen to occur with a probability proportional to its rate constant $k_{event}$, determined by another random number $u_2$. Subsequently, the system transit to the new configuration, and the available event list is updated accordingly. These steps are illustrated as a flow chart in Fig. 1. $k_{event}$ is a site-specific parameter that depends on the energy profile of the site (see Section III).

## III. `MOSP` implementation

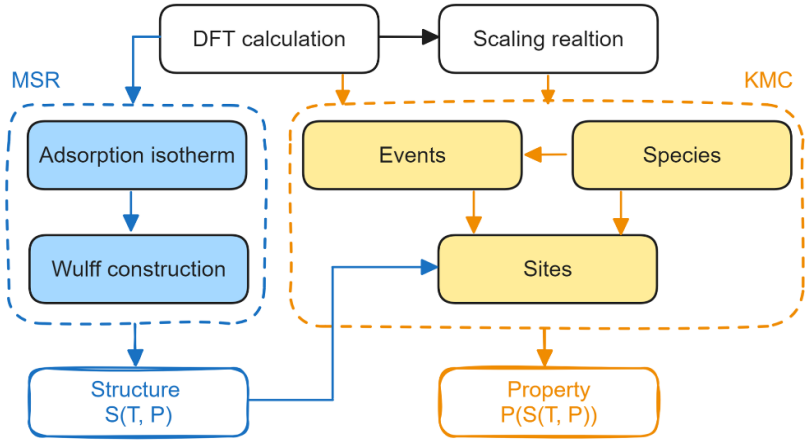

Fig. 2. Workflow of MOSP package

The graphical user interface (GUI) of `MOSP` is built using the `python` library `wxpython`.[37] All required and optional computational parameters are set through the GUI. Furthermore, a visual panel is included for the nanoparticle structure visualization (using `pyOpenGL`[38]) and the KMC simulation results representation (using `pandas`[39] and `matplotlib`[40]). The workflow of `MOSP` is illustrated in Fig. 2. Based on first-principles data, the equilibrium structure of NPs at a given reactive atmosphere $S(T, P)$ can be constructed by the MSR module. This structure can be loaded as an input of the KMC module to generate a set of sites. Then, KMC simulation can be carried out after setting the reaction intermediate (Species) and elementary reactions (Events) to evaluate the corresponding property $P(S(T, P), T, P)$, including coverages and turnover frequencies (TOFs).

While the combined use of MSR and KMC is suggested to comprehend the dynamic structure-activity relationship, both modules can also be employed independently. The MSR-constructed structure can export directly, whereas KMC module can load an existing structure $S$ in the XYZ format to derive $P(S, T, P)$.

**A. Model definition**

In MOSP, nanoparticles are modeled as a collection of discrete sites. Each site is characterized by three fundamental attributes: element, position, and occupation value. If the particle is constructed through MSR, an additional attribute, type, is assigned to each site. This attribute specifies the site's specific type, such as facet, edge, corner, subsurface, or bulk. In the KMC module, neighbor lists for each site are generated based on bond length, followed by the assignment of coordination number (CN) and generalized coordination number[41, 42] (GCN). GCN is utilized as a descriptor of adsorption energies, facilitating the mapping of site-specific event rates in KMC simulations.[43, 44] More descriptors may be supported in future releases.

Species are classified into two groups: adsorbates and products. Each species is identified by a unique name and internally mapped to an integer upon definition, serving as an occupation value option for sites. For products, this information is sufficient as they are only used as labels during data post-processing for TOF calculations. For adsorbates, it is necessary to define the scaling relationship parameters that link their adsorption energies to their GCNs, along with the parameters required for calculating the event rates in which they are involved. Additionally, the lateral interactions between adsorbates, denoted as $w$, need to be set in an $n \times n$ symmetric matrix, where $n$ is the number of adsorbates.

Each event evolves one or two neighboring surface sites. Defining an event requires specifying the event type and the occupation values of the involved sites before and after the event. Event types can be categorized as adsorption, desorption, diffusion, or reaction, which will determine the rate constant expressions used for the event.

For adsorption, the rate constants are obtained from collision theory (CT), assuming zero activation energy[45]. The adsorption rate constant for species $i$ is given by:

$$k_{ads,i} = \frac{s_{0,i} A_{at} P_i}{\sqrt{2\pi M_i k_B T}} \quad (8)$$

where $s_{0,i}$ is the sticking coefficient, and $M_i$ is the molecular mass of species $i$. The

corresponding desorption rate is defined to satisfy the thermodynamic equilibrium:

$$k_{des,i} = \frac{k_{ads,i}}{K_i P_i} \qquad (9)$$

Here, $K_i$ is the equilibrium constant as defined in Eq. 4, with the adsorption energy here being a site-specific value determined linearly by the GCN of the involved site.

For reaction and diffusion, the rate constants are derived from transition state theory (TST)[46]:

$$k = \frac{k_B T}{h} \exp\left(-\frac{E_a}{k_B T}\right) \qquad (10)$$

where $h$ is the Plank constat, and $E_a$ is the energy barrier. The diffusion barrier $E_a^{diff}$ is specified in the definition of the species involved, while the reaction barrier $E_a^{rec}$ is determined using the Brønsted-Evans-Polanyi (BEP) relation.[47, 48]

## B. Inputs for MOSP

To set up a simulation, the required inputs are listed in Table I. The inputs can be saved and loaded in human-readable JSON format through GUI. Sample inputs are provided in the `examples` folder of our GitHub repository (see the Data Availability).

**Table I.** List of inputs required by MOSP, including corresponding symbols for inputs mentioned in Section II and III, and units for non-dimensionless quantities.

| Module | Input | | Symbol (Units) |
|---|---|---|---|
| Common | Element | | |
| | Crystal Structure | | |
| | Lattice constant | | (Å) |
| | Pressure | | $P$ ($Pa$) |
| | Temperature | | $T$ ($K$) |
| MSR | Radius of particle | | (Å) |
| | Gas | Name | |
| | | Partial Pressure ratio | $P_i$ |
| | | Standard entropy | $S^0_{gas,i}$ ($eV/K$) |
| | | Adsorption type | $\alpha_i$ |
| | Face | Index | $hkl$ |
| | | Surface tension | $\gamma_{hkl}$ ($eV/Å^2$) |
| | | Adsorption energies | $E^{hkl}$ ($eV$) |
| | | Entropies of adsorbates | $S^{hkl}_{ads,i}$ ($eV/K$) |
| | | Lateral interaction between adsorbates[a)] | $z \times w^{hkl}$ ($eV$) |
| KMC | Number of KMC steps | | |
| | Record interval | | |
| | Adsorbate | Name | |
| | | Is_twosite[b)] | |
| | | Scaling relation parameters[c)] | |

|         |                                               |                      |
|---------|-----------------------------------------------|----------------------|
|         | Molecular Mass[d]                             |                      |
|         | Sticking coefficient[d]                       | $s_{0,i}$            |
|         | Partial pressure ratio of gaseous specie[d]   | $P_i\ (Pa)$          |
|         | Standard entropy of gaseous speciec[d]        | $S^0_{gas,i}\ (eV/K)$ |
|         | Entropy of the adsorbate[d]                   | $S_{ads,i}\ (eV/K)$  |
|         | Diffusion barrier[d]                          | $E_a^{diff}\ (eV)$   |
|         | Lateral interaction between adsorbates        | $\boldsymbol{w}\ (eV)$ |
| Product | Name                                          |                      |
| Event   | Name                                          |                      |
|         | Is_twosite[e]                                 |                      |
|         | Event type                                    |                      |
|         | Reactants                                     |                      |
|         | Products                                      |                      |
|         | BEP relation parameters[d]                    |                      |

[a] That is, $z \times w$ in Eq. 1, 2. The total lateral interaction at one monolayer adsorption[16].

[b] A Boolean value, true if the adsorbate occupies two adsorption sites.

[c] $E_{ads} = k_1 GCN_1 + b$ for a single-site species, $E_{ads} = k_1 GCN_1 + k_2 GCN_2 + b$ for a two-site species.

[d] Optional inputs.

[e] A Boolean value, true if the event involves two sites.

## IV. Examples

### A. CO oxidation over Pt nanoparticle

In our previous studies,[21, 22, 49] the oxidation of CO over Pt NPs has been thoroughly investigated using MSR and KMC. Here, we use this reaction as an example to illustrate the workflow and performance of `MOSP`.

**Table II.** Data required for MSR of Pt NPs in CO-oxidation atmosphere: surface tension under vacuum $\gamma_{hkl}\ (eV/Å^2)$, adsorption energies $(eV)$ and lateral interactions $(eV)$.[21]

|          | $\gamma_{hkl}$ | $E_{CO}$ | $E_O$   | $z \times w_{CO-CO}$ | $z \times w_{O-O}$ | $z \times w_{CO-O}$ |
|----------|----------------|----------|---------|----------------------|--------------------|---------------------|
| Pt (100) | 0.112          | -1.703   | -1.026  | -0.688               | -0.516             | -0.548              |
| Pt (110) | 0.113          | -1.865   | -1.153  | -0.384               | -0.448             | -0.426              |
| Pt (111) | 0.087          | -1.494   | -0.982  | -1.188               | -1.056             | -0.756              |

First, the MSR module was used to construct the equilibrium structures of Pt NPs with a radius of 2.5 nm under varying pressure, maintaining a fixed partial pressure ratio ($P_{CO}:P_{O_2} = 3:2$) at 800 K. Fig. 3(a) displays a snapshot of the `MOSP` GUI after the completion of the MSR at 500 Pa. The left panel serves as the input interface, where the lattice information of Pt and reaction conditions are specified. The partial pressure ratio, standard gas entropy $S_{gas}^0$, and adsorption types of CO and $O_2$ are then defined. Furthermore, the adsorption energies, entropy, and lateral interactions on each crystal facet are assigned, which can be obtained through Density Functional Theory (DFT) calculations. Further details and specific parameters are available in Table II and our previous work.[21] The right panel contains a visual panel and a log table. In the visual panel, the structure of the constructed particle is demonstrated, colored according to the surface site type (green: (100), red: (110), blue: (111), pale yellow: edge, dark yellow: corner). Fig. 3(b) illustrates the fractions of different surface site types under varying pressures. As pressure increases, the fractions of the (111) facet decreases, while the fractions of the (110) facet increases. Consequently, the shape of the Pt NP changes from a truncated octahedron to a quasi-rhombic dodecahedron. Three typical structures at 0, 1000 and 2000 Pa are displayed in Fig. 3(c).

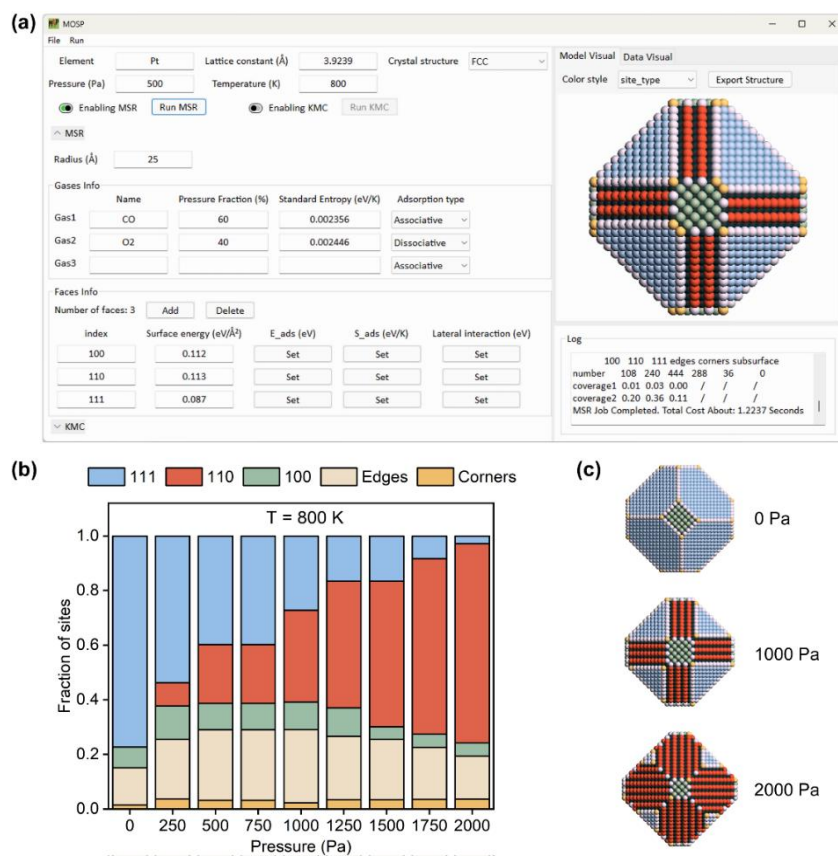

**Fig. 3.** (a) Inputs and outputs of the MSR module for a 5nm Pt NP in CO-oxidation atmospheres at 500 Pa. (b) Surface site type fractions of the Pt NPs at different pressures ($T = 800\ K$, $P_{CO}:P_{O2} = 3:2$). (c) Typical constructed structures at 0, 1000, and 2000 Pa.

Next, the KMC module was used to investigate the catalytic properties of the constructed particle at 500 Pa. Fig. 4(a) shows the input panel of KMC module. For CO oxidation on Pt NPs, the KMC model includes two adsorbates (CO* and O*), one product (CO2) and seven events (associative adsorption and desorption of CO, dissociative adsorption and desorption of $O_2$, diffusion of CO and O, as well as CO oxidation). The KMC simulation ran for 10 million steps with a recording interval of 50 thousand steps. After the simulations, the particle was visualized with a colormap of GCN (Fig. 3b) and normalized site-specific TOF (Fig. 3c). In this condition, edges and corners make the major contribution to total activity, while the facets are inactive. Fig. 5(d) shows the coverage of CO and O as a function of time. The TOF trend is calculated as the reaction rate per surface site over ten evenly spaced intervals. In this example, the system has reached equilibrium after 10 million KMC steps. The results from the last one million steps were used for statistics, with CO and O coverages at 0.01 and 0.11, respectively, and a TOF of $8.33 \times 10^5 \ s^{-1} site^{-1}$.

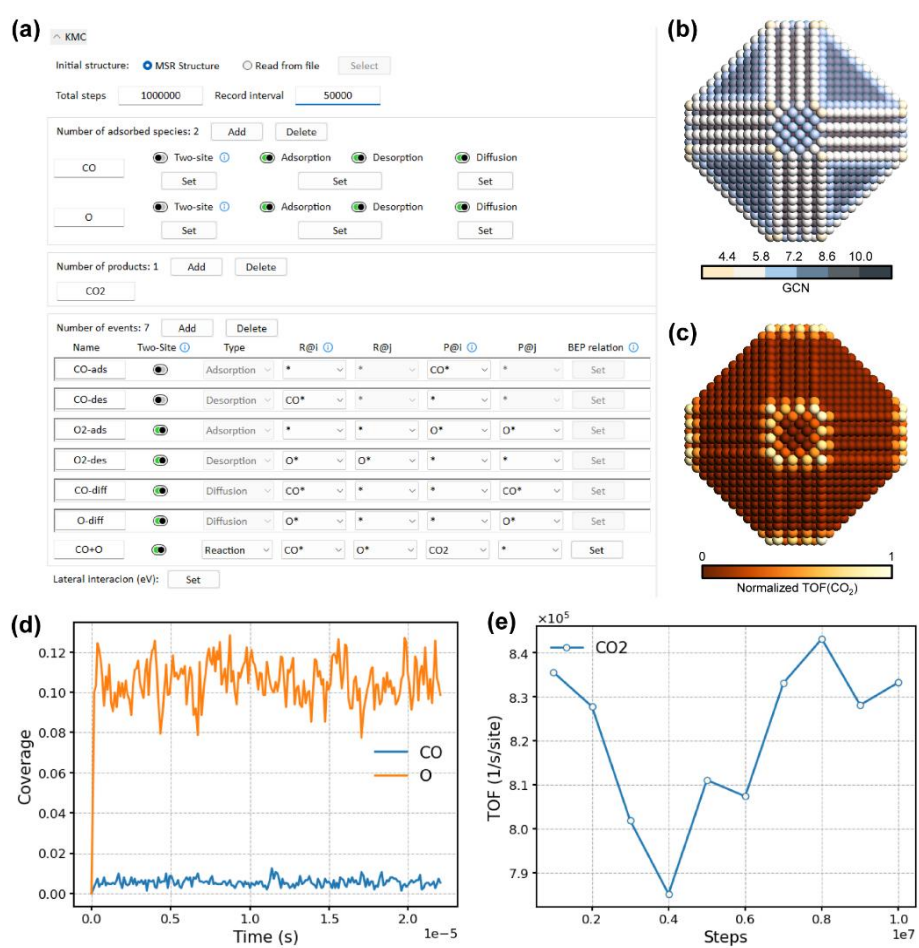

**Fig. 4.** (**a**) Inputs of the KMC Module for CO oxidation on Pt NPs. Model visualization with atoms colored according to the GCN (**b**) and normalized site-specific TOF (**c**). Data visualization of the coverage trends (**d**) and TOF trend (**e**). The results are obtained at 800 K with a total pressure of 500 Pa ($P_{CO}:P_{O_2} = 3:2$).

To investigate the performance dependence on the number of sites. The CPU time required to execute MSR and one million KMC steps for particles of varying sizes was tested on a benchmark system with a 2.1 GHz Intel Core i7 processor and 16 GB RAM (Fig. 5). The data points were fitted with a power function ($t \propto N_{atoms}^a$) and the fitted functions are depicted as dashed lines. Typically, the power for MSR and KMC (per Million steps) is 2.4 and 0.93 respectively.

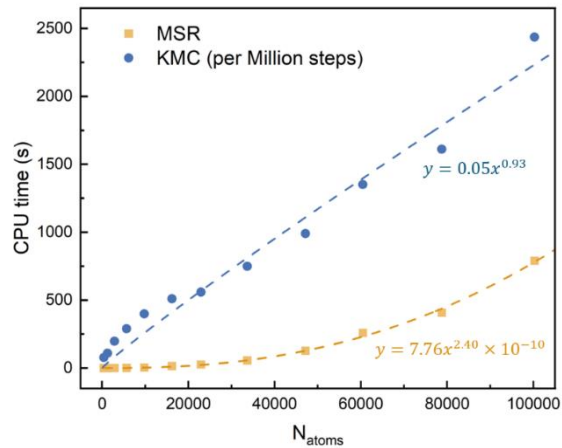

**Fig. 5.** Single-core CPU times required to execute MSR and one million KMC steps for CO oxidation on Pt NPs with varying numbers of atoms. The benchmark was conducted on a 2.1 GHz Intel Core i7 processor with 16 GB of RAM.

**B. Water-Gas Shift Reaction over Cu nanoparticle**

To demonstrate the scalability of KMC module, we consider the water-gas shift (WGS) reaction over Cu NPs as proposed by Xu[44]. The WGS reaction on a Cu metal surface follows the carboxyl mechanism, involving the following events:

$$CO(g) + * \rightarrow CO^* \tag{R1}$$
$$CO^* \rightarrow CO(g) + * \tag{R2}$$
$$H_2O(g) + * \rightarrow H_2O^* \tag{R3}$$
$$H_2O^* \rightarrow H_2O(g) + * \tag{R4}$$
$$H^* + H^* \rightarrow H_2(g) + 2* \tag{R5}$$
$$H_2O^* + * \rightarrow OH^* + H^* \tag{R6}$$
$$CO^* + OH^* \rightarrow COOH^* + * \tag{R7}$$
$$COOH^* \rightarrow CO_2(g) + H^* \tag{R8}$$
$$COOH^* + OH^* \rightarrow CO_2(g) + H_2O^* \tag{R9}$$

Here, an asterisk * stands for an empty surface site, and the adsorbates are labelled with superscript asterisks. For simplicity, $CO_2$ is considered to desorb immediately after formation. The model includes five adsorbates (CO*, H$_2$O*, OH*, H*, and COOH*), two products (CO$_2$ and H$_2$), and fourteen events (R1-R9 and the diffusion of the five adsorbates).

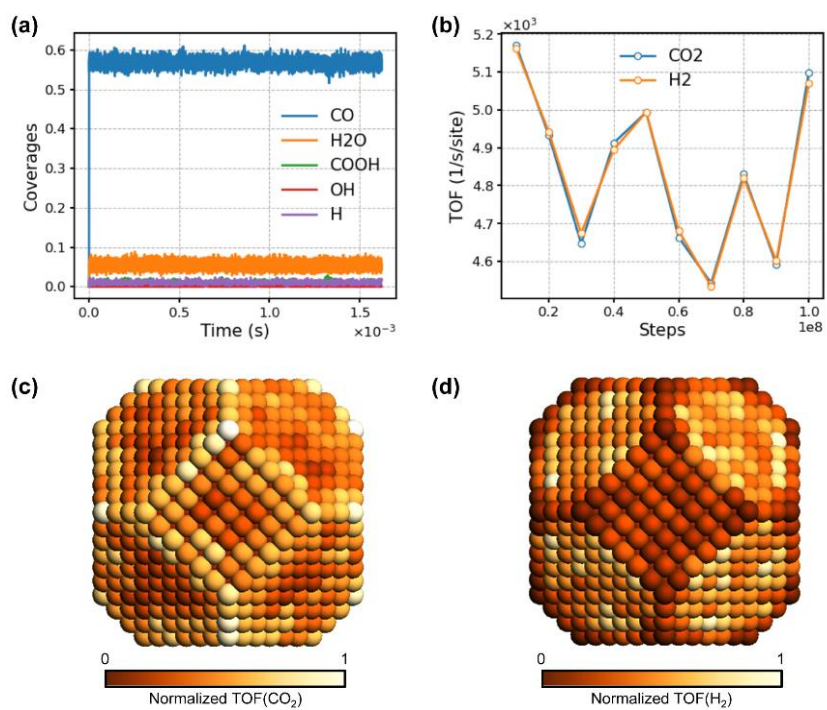

**Fig. 6.** Data visualization of the coverage trends (**a**) and TOF trend (**b**). Model visualization with Cu atoms colored according to the normalized site-specific TOF($CO_2$) (**c**) and TOF($H_2$) (**d**). The results are obtained at 650 K with a total pressure of 4000 Pa ($P_{CO}:P_{H_2O} = 1:1$).

The KMC simulation was conducted using a truncated octahedron Cu NP containing 2190 atoms at 650 K and a total pressure of 4000 Pa ($P_{CO}:P_{H_2O} = 1:1$). The simulation ran for 100 million steps with a recording interval of 50 thousand steps. As in the previous example, the coverages and TOFs trends, as well as the normalized site-specific TOFs, are visualized in Fig. 6. In this model, the most abundant adsorbate is CO, with a steady-state coverage of 0.67. The next most abundant is $H_2O$, with a coverage of 0.05, while the coverages of other adsorbates are all below 0.01 (Fig. 6(a)). The TOFs of $CO_2$ and $H_2$ on the whole particle are nearly the same, at $5.10 \times 10^3$ and $5.07 \times 10^3 \ s^{-1}site^{-1}$ respectively (Fig. 6(b)). However, the active sites for $CO_2$ and $H_2$ formation differ. As shown in Fig. 6(c) and 6(d), low-coordination sites are favorable for $CO_2$ formation, while high-coordination ones are favorable for $H_2$ formation. This example effectively demonstrates the kinetic coupling in NPs at the atomic level.

## V. Conclusion

In summary, we have presented the open-source package MOSP, which enables the simulation of structures of metal NPs under operando conditions and the estimation of their catalytic performances. The particle reconstruction, customized KMC simulation of reaction, along with the post-processing and visualization of the simulation results, are readily implemented in `MOSP` through an interactive GUI. Its capabilities have been demonstrated with illustrative examples. `MOSP` is designed for use on standard laptops, providing researchers with an efficient tool to predict the structure and catalytic properties of NPs in the reaction atmosphere, aiding in the screening of optimal reaction conditions.

## ACKNOWLEDGMENTS

This work was supported by National Key R&D Program of China (2023YFA1506903), National Natural Science Foundation of China (12174408), Natural Science Foundation of Shanghai (22JC1404200), Shanghai Municipal Science and Technology Major Project, and the Key Research Program of Frontier Sciences, CAS, Grant No. ZDBS-LY-7012.

## DATA AVAILABILITY

The source code is available at https://github.com/mosp-catalysis/MOSP.